\documentclass[reprint, amsmath,amssymb,aps, longbibliography]{revtex4-1}%
\usepackage{graphicx, graphics}
\usepackage{wrapfig}
\usepackage[centerlast]{caption}
\usepackage{subcaption}
\usepackage{natbib} 
\usepackage{multirow}
\usepackage{float}
\usepackage{hyperref}
\usepackage{psfrag}
\usepackage{amsmath}
\usepackage{amssymb}
\usepackage{gensymb}
\usepackage{here}
 \usepackage[framemethod=tikz]{mdframed}

\definecolor{mycolor}{rgb}{0.122, 0.435, 0.698}

\usepackage{fancyhdr}
 
\newmdenv[innerlinewidth=0pt, 
linecolor=black,innerleftmargin=6pt,
innerrightmargin=6pt,innertopmargin=6pt,innerbottommargin=6pt]{mybox}

\begin{document}
	
\title{Examining and contrasting the cognitive activities engaged in undergraduate research experiences and lab courses}

\author{N.G. Holmes}
\email[]{ngholmes@stanford.edu}
\affiliation{Department of Physics, Stanford University, Stanford, CA}
\author{Carl E. Wieman}
\affiliation{Department of Physics, Stanford University, Stanford, CA}
\affiliation{Graduate School of Education, Stanford University, Stanford, CA}

\date{\today}

\begin{abstract}
While the positive outcomes of undergraduate research experiences (UREs) have been extensively categorized, the mechanisms for those outcomes are less understood. Through lightly structured focus group interviews, we have extracted the cognitive tasks that students identify as engaging in during their UREs. We also use their many comparative statements about their coursework, especially lab courses, to evaluate their experimental physics-related cognitive tasks in those environments. We find there are a number of cognitive tasks consistently encountered in physics UREs that are present in most experimental research. These are seldom encountered in lab or lecture courses, with some notable exceptions. Having time to reflect and fix or revise, and having a sense of autonomy, were both repeatedly cited as key enablers of the benefits of UREs. We also identify tasks encountered in actual experimental research that are not encountered in UREs. We use these findings to identify opportunities for better integration of the cognitive tasks in UREs and lab courses, as well as discussing the barriers that exist. This work responds to extensive calls for science education to better develop students' scientific skills and practices, as well as calls to expose more students to scientific research.
\end{abstract}

\maketitle
\section{Introduction}
A recent focus of science education has been on building scientific skills and practices, beyond content mastery \cite{Quinn2012, AAPT14}. Instructional laboratory courses (labs) and undergraduate research experiences (UREs) are key platforms through which undergraduate students can experience these practices in a ``hands-on/minds-on" way. The American Physical Society, in particular, has recently called upon higher education institutions to increase student access to UREs \cite{APS}, citing the positive outcomes of such experiences. Given the costs of these experiences (as well as other resource requirements), providing UREs to all students seems logistically impractical, especially given that it has been estimated that only 50\% of undergraduate science students are currently accessing UREs \cite{Russell2007}. Many attempts have been made, therefore, to incorporate aspects of UREs into course-based activities \cite[e.g.][]{Brownell2015, Auchincloss2014, Brownell2011}, despite a lack of evidence towards the particular mechanisms that are responsible for the observed outcomes \cite{Laursen2015}. This has emphasized the need to better understand what about UREs leads to positive student outcomes.

To this aim, this paper aims to explore the activities in which students engage during their UREs, using the cognitive tasks of experimental physics \cite{Wieman2015} as a lens for that exploration (Fig. \ref{fig:tasks}). We also begin to identify the uniqueness of these tasks in UREs by comparing them to the tasks in which students engage during instructional labs and other coursework. We use the comparisons to suggest what aspects of UREs could best be incorporated into course activities. 

\begin{figure}
\begin{center}
\fbox{\begin{minipage}{23em}
Cognitive Task Analysis Elements
\begin{enumerate}
\item Establishing research goals
\item Defining criteria for suitable evidence
\item Determining feasibility of experiment
\item Experimental design
\item Construction and testing of apparatus/code
\item Analyzing data
\item Evaluating results and analyzing implications
\item Presenting the work
\end{enumerate}
\end{minipage}}
\end{center}
\caption{List of experimental physics cognitive task analysis (EPCTA) elements from \cite{Wieman2015}. Each task element contains a number of sub-tasks and there is extensive iteration between tasks and sub-tasks, as well as cycling back to earlier elements. Note that element seven combines two tasks from the original work.}
\label{fig:tasks}
\end{figure}

\begin{table*}
\caption{Demographics of the sample of students included in the interviews.} \label{tab:participants}
\begin{center}
\begin{tabular}{lp{3cm}p{4cm}}
\hline\hline
Curricular level & Number interviewed & Number in URE\\
\hline
Rising sophomores (completed freshman/first year) & 16 & 19\\
Rising juniors (completed sophomore/second year) & 10 & 19\\
Rising seniors (completed junior/third year) & 4 & 12 \\
\hline
Number of faculty members of students interviewed & 21 & 24 \\
Number of women & 9 (30\%) & 13\\
\hline\hline
\end{tabular}
\end{center}
\end{table*}%

Prior research probing the outcomes of UREs have highlighted clarifying career aspirations, understanding the scientific research process, learning how to think like a scientist, broadening disciplinary content knowledge, and developing self-confidence and self-efficacy \cite{Sadler2009, Thiry2011, Thiry2012,Hunter2007, Seymour2004, Lopatto2004}. While these elements emerge as common themes in students' reported gains, the extensive variability in each student's experience relates to large variability in an individual student's gains \cite{Hanauer2012}. Developing experimentation skills and ways of thinking has been studied a limited amount, notably in the work of Hunter et al \cite{Hunter2007}.
Here we are building on this past work by digging more deeply into the particular skills and activities in which students engage in UREs. We then examine how well these overlap with the full range of cognitive tasks involved in carrying out actual experimental physics research, and contrast this to students' course-based experiences.

This examination uses the recent analysis of the cognitive tasks used by physicists to successfully carry out an experiment from conception to completion \cite{Wieman2015}. Wieman defines the Experimental Physics Cognitive Task Analysis (EPCTA) items as the ``mental tasks or types of thinking (`cognitive task analysis') associated with a physicist doing tabletop experimental research" \cite[p. 1][]{Wieman2015}. Figure \ref{fig:tasks} lists the overarching cognitive tasks, each of which involves a number of sub-tasks. Though the list suggests a linear process, Wieman \cite{Wieman2015} specifies that moving through the EPCTA often occurs in a highly iterative and cyclic process. While these tasks were developed in the context of experimental physics, many of these elements are fundamental skills and practices that will benefit students in a variety of contexts. 

As lab courses involve carrying out experiments, it is natural to consider the extent to which they emulate actual research. It should be acknowledged, however, that this is not necessarily the sole, or even primary, goal of all lab courses. The goals of lab courses have been highly debated and without consensus for many years \cite{Hofstein1982, Hofstein2004, ALR}. There is suggestive evidence, however, that traditionally-taught lab courses provide little added value for developing or improving content understanding beyond lectures or tutorials \cite{WiemanAJP, WiemanPERC15, Hofstein2004, Sere2002}. The hands-on experimentation in lab courses provides a unique opportunity, however, for developing students' scientific and experimentation skills. In addition, limited resources provide challenges to consistently offering research opportunities to undergraduate students, and so it is important to examine the extent to which lab courses can compensate for these limitations. Furthermore, this analysis will provide insight into the uniqueness of UREs at the cognitive level, to shed light on what may be lost in replacing UREs with course-based activities.

Here we are considering the traditional lab course and not course-based undergraduate research experiences (CUREs) \cite[e.g.][]{Makarevitch2015, Shapiro2015, Brownell2011, Brownell2015, Auchincloss2014}. While they offer an interesting solution to these issues, none of our study population had experience in CUREs and so we did not consider them in this work.

The research questions we are trying to answer are: 
\begin{itemize}
\item What cognitive tasks do students engage in during UREs, as contrasted to their experiences in course-based activities (especially through lab courses)?
\item What barriers limit this engagement in UREs, coursework, and lab coursework?
\item In what ways could coursework and lab coursework better incorporate these tasks (with the aim of potentially facilitating larger scale participation and/or richer UREs)?
\end{itemize}

\section{Methods}

Nine hour-long focus group interviews were conducted with summer URE students in the physics department at an elite university. All 51 students conducting research in the department were invited to participate through several email notices and short recruitment presentations during professional development events. Students signed up to participate after each such call and 32 of the 51 individual students participated in an interview by the end of the program. 

The physics department at this institution guarantees funding for one summer of research for every physics major and all students that apply to work each summer are accepted. This provides a unique measurement opportunity in that our sample provides the perspective of students at a range of seniorities and class standings (Table \ref{tab:participants}). Most studies of undergraduate research experiences are limited to juniors or seniors (upper-division students) with high grade-point averages \cite{Russell2007}. It must still be recognized, however, that the physics majors in our sample are somewhat unusual; for example, several had had research experience while in secondary school.

\subsection{Interviews}

\begin{figure}
\vspace{-2em}
 \begin{mybox}
\begin{enumerate}
\item What year are you and what is your background experience in research?
\item Why did you sign up to do an URE project this summer?
\item How are you enjoying your research so far? Can you give a little background about why this project is interesting to you?
\item What has been most enjoyable or rewarding to you about research so far?
\item What has been most disappointing?
\item What has been most different from your expectations (good or bad)?
\item What are the most important things you have learned? 
\item What are some things you wish you had known before starting out your research project?
\item What are some things that you did know starting out your research project that you think have helped you succeed?
\item Have your views about how research works and how much you enjoy it changed, and if so how and why?
\item Which scheduled development sessions did you find most/least useful/enjoyable?
\item Tell me about your progress on the goals that you set at the beginning of the summer.
\item How do you know that you have been making progress towards those goals?
\item Tell me about how you came up with your goals. Who was responsible for deciding on the goals?
\item In what ways has your experience in lab courses been similar to, or different from, your research experience?
\item In what ways has your experience in lab courses prepared you to do your research project?
\item Is there anything else you would like to tell me about your research experience?
\end{enumerate}
\end{mybox}
\caption{List of questions used in each interview, though the order of questions varied.}\label{fig:Qs}
\vspace{-2em}
\end{figure}

Interviews were conducted across eight weeks of the 8-10 week summer research program. The number of participants in a given interview varied from two to eight students. In addition, 12 students were interviewed twice. No two interviews, however, were made up of the same set of students. That is, the composition of each interview differed, so students were interacting with different peers and, therefore, different ideas. In addition, students interviewed twice were interviewed towards the start and end of the summer, so the topics they focused on naturally shifted.

Interviews were semi-structured. There was a fixed set of targeted questions to be asked during each interview (Fig. \ref{fig:Qs}). As can be seen from the list, most questions were explicitly centered on students' URE. The order of the questions differed and particular questions were probed more deeply depending on student answers. In addition, since there were multiple participants in each session, ideas evolved and expanded as students elaborated on or provided counter arguments to their peers' statements. All questions were discussed in each of the interviews. The interviewer asked every student to answer the first couple of questions by going around the table. For subsequent questions, students were participating freely in the discussion. While different students participated to varying degrees, the interviewer attempted to encourage all students to participate throughout, and in no interview did a single student (or two) dominate the discussion, and in all interviews all students contributed to the discussion at some point.

\subsection{Courses}

At the introductory level, students enrolled in three 10-week physics courses, each of which involved optional associated lab courses. The lecture components of the introductory courses included some interactive engagement, but, in general, their courses were relatively traditional in their pedagogical approaches. The associated lab courses, however, varied in their pedagogical designs and learning goals. A \emph{design} lab course engaged students in a single, extended, student-designed experiment. The other two courses were \emph{structured} lab courses that engaged students in highly structured lab exercises that changed weekly. The structured labs aimed to reinforce or develop concepts related to the lecture course content, with some focus on exposing students to a variety of equipment or teaching data analysis and statistics. All students had taken at least one structured lab course and about 90\% of students interviewed had taken the introductory design lab course. 

A few advanced students in the interviews discussed a senior-level project lab course, where students work in groups to design and carry out their own experimental project in low-temperature physics. This course will be discussed in the context of design lab courses. An upper division electronics course involved some troubleshooting of circuits and experimental set ups, but was otherwise relatively structured and traditional. This course is therefore discussed in the context of the structured lab courses.

\subsection{Interview analysis}
The interviews were first analyzed for emergent themes. A number of the themes that emerged were consistent with previous research on outcomes of UREs (e.g. learning about career choices, the process of science, the life of academics). Evidence of this consistency can be found in the sample quotes embedded. 

Our questions did, however, elicit extensive discussions that probed comparisons between the URE and their prior lecture and lab courses. Primarily, these responses were related to what kinds of activities students were and were not doing in each context, what affordances one provided over the other, and students' opinions of these activities. Some of the most prominent topics were troubleshooting, experimental design, opportunities and time for reflection and iteration, the lack of a single or clear correct answer to a problem, autonomy, and comparisons with course work. 
The EPCTA provided an appropriate framework for characterizing students' cognitive processes.

Interviews were coded based on the EPCTA. Instances of a particular cognitive task were categorized (or coded) for whether students were referring to the task in research, lecture courses, or the structured or design lab courses. The context of the discussion was also recorded as to whether students explicitly discussed that they were doing it, were not doing it, or whether discussion was mixed. From those notes, the coded discussions were further refined for common themes.

Our methodology places limitations on the analysis we perform and the conclusions we can draw. We cannot use the frequency of particular statements or cognitive tasks in a particular interview to quantitatively represent importance or value. That is, a cognitive task being discussed more times or for a longer period of time in one interview is more representative of the nature of the interview than of the value students placed on that idea. We also cannot decisively say that because a code did not come up in an interview that it was not relevant in their research or classwork. The use of repeated interviews, however, does provide some quantitative insight. Namely, importance can be inferred if a task element was discussed in multiple interviews. In contrast, a lack of relative value (compared with the other tasks) can be inferred if a task element was discussed in few interviews or if many interviews explicitly identified not engaging in that task. It would be desirable, therefore, to follow this work with targeted interviews or surveys with different samples of students in different contexts to better establish the generalizability and consistency of the results.

\section{Results}

\begin{figure*}
\begin{center}
\includegraphics[width=\textwidth]{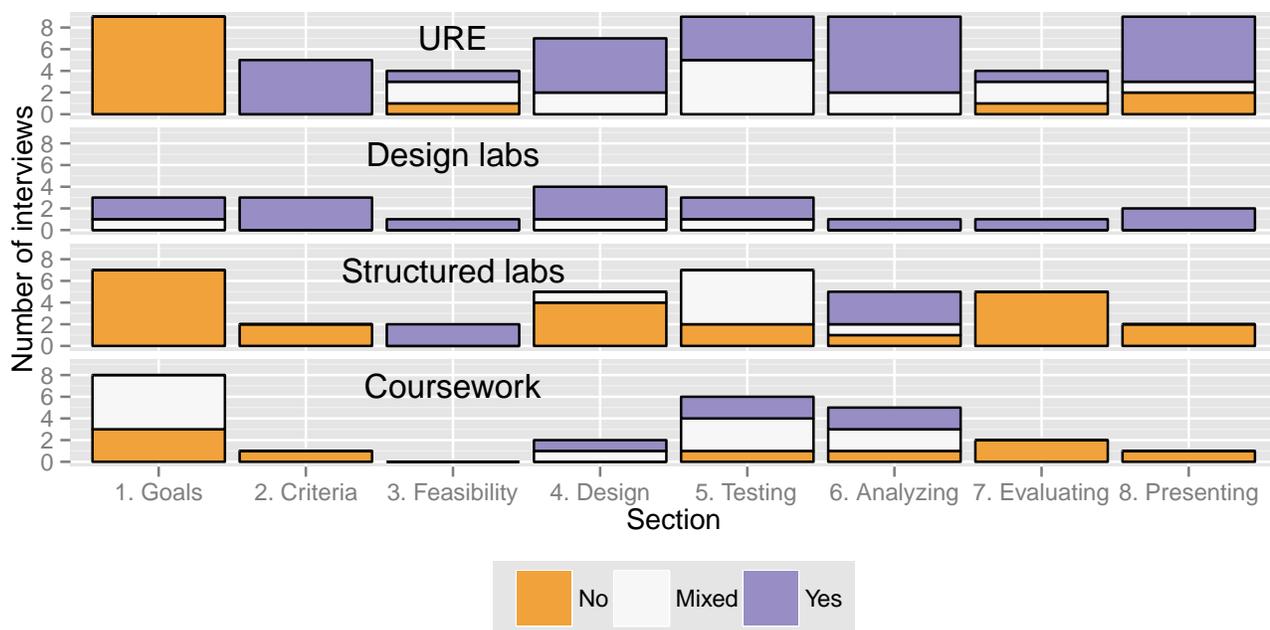}
\caption{Fraction of interviews in which the EPCTA elements were discussed in the context of URE, classwork, or the structured or design lab courses. Comments were categorized as Yes (students were performing this task), No (students were not performing this task), or mixed (some students were and others were not).}
\label{fig:Graphs}
\end{center}
\end{figure*}

In what follows, we describe the details of discussion surrounding each of the cognitive tasks independently. The summary of the number of interviews discussing each cognitive task element, as well as the context of that discussion, can be found in Fig. \ref{fig:Graphs}. We provide sample quotes throughout the text. These use speaker identifiers \emph{NH} to reflect the initial of the interviewer and \emph{S} to reflect the student (with \emph{S1} and \emph{S2} and so on to distinguish multiple speakers).

\subsection{Establishing research goals}

Students unanimously reported that the goals of their URE projects were provided to them by their research mentor. Related to predicting whether the goal was ahead of current knowledge, most of the discussion centered on whether the goal was ahead of \emph{their} knowledge. That is, students generally felt that their classwork had not prepared them to deal with the content in their projects. Students did identify that their courses laid groundwork for the research content. This was either in core courses that introduced basic terminology or through broad perspective courses that introduce current research topics.

Few students discussed deciding whether the project was feasible, presumably because the supervisor had already chosen the project. In two interviews, students discussed some lack of knowledge regarding goal setting. For example, they wondered how these projects come to fruition: how do we know this is worth doing? Another student wanted to know more about how to apply information from research papers to solve new problems. 

Students also unanimously reported that the specific experimentation goals were always given in their structured lab courses. Students described that, often in labs, the goal was to ``obtain this number" (e.g. gravitational acceleration as 9.8 m/s$^2$). They directly contrasted this to their URE, where the goal was rarely to obtain a single number, and never to obtain a given number. In the design lab course, in contrast, students described having some control over the research question, though the space was constrained to a particular mechanics context (at the intro level) or low-temperature physics (at the advanced level). Only one student differentiated the URE as uniquely producing new knowledge.

\subsection{Defining criteria for suitable evidence}

In UREs, the notion of identifying which variables were important arose in the context of the importance of understanding every process and piece of their project or apparatus. This topic was picked up in one focus group with three different students contributing to the discussion, all of whom were upper-division students who had completed more than one summer URE. One student explicated how the fact that there was no `right' answer required them to carefully evaluate all the relevant variables. Related to the design lab courses, students similarly described evaluating the relevant variables and how to measure or control them, though they described the process as much simpler and more constrained. In the URE, the students found that the sheer number of relevant variables required them to integrate many different areas of physics. This was contrasted to the lack of integration of different topics of physics in their typical homework assignments. 

A student also expressed surprise at realizing different researchers took different approaches to model the same system, which reinforced the need to carefully justify their decisions:

\begin{mybox}
{\noindent \emph{Student (S)}: ``When you're building this trap and ... you're supposed to try to levitate this tiny bead. And the difficult part is that even when you have all your optics perfectly aligned and everything, there are still so many parameters that go into whether your trap can trap or not. And figuring that out is such a frustrating and long process because basically you're just bashing at a black box. ... But over time you sort of build this sensitive understanding of each of the different components of this black box. And at the end, when you understand the big picture, I think that's the most satisfying."
}
\end{mybox}

Students briefly touched on the notion of determining what evidence would be convincing and for whom. A student reflected on how the goal of a previous URE project shifted significantly when they did preliminary measurements and the effect was not as strong as expected. Another student described how researchers need to ``see stuff to show that it's there... no argument matters until you have a plot of a 5-sigma effect." This student described learning about how the goal of research was to convince a ``highly skeptical community" of researchers. This was explicitly contrasted to their classwork, where there was no such element of argumentation. 

Students did not specifically discuss activities related to identifying necessary controls or checks. Students also did not discuss evaluating criteria with regards to their structured lab courses, because, as they described, they simply ``followed the instructions and got to the end, and turned it in."

\subsection{Determining feasibility of experiment}

The main topic of discussion about evaluating feasibility, in all contexts, was that of time. Students said that having a fixed amount of time in their URE caused their goals to change, especially because timescales for completion were much longer than they expected:

\begin{mybox}
{
\noindent \emph{S}: ``There was a big sort of shift for me about two days ago, but it was sort of coming on for the past few weeks. Originally I was planning on getting a lot more done this summer and actually doing the trapping of atoms, but now I'm realizing I'm just going to have built a vacuum chamber and that's not such a terrible thing to have done. Like, such a small amount of stuff I've accomplished in, you know, 2 and a half months."

\noindent \emph{Interviewer (NH) asks what caused the shift.}

\noindent \emph{S}: ``Well, at the very beginning of the summer ... [my mentor] said we'd probably have time in 10 weeks to do this but everything always takes longer than you think it will and you have to order parts and those take a while to get there and you have to take them to the shop and the shop takes 4 days to do whatever you need done."
}
\end{mybox}

Regarding contingency plans, some students identified that they had back-up project goals if things did not go as planned. One student emphasized the importance of having several actionable goals so that, even if large roadblocks are met, the student can still make progress on some pieces. 

There was little discussion of feasibility in the context of labs or coursework. The design lab course was highlighted in that it gave students the flexibility to shift their goal if things were not working out as planned, similar to UREs. Historical experiments included in the structured labs (such as a Millikan oil-drop experiment) were described as providing appreciation for how difficult these classic experiments had been, though the results are fundamental to our knowledge now.
 
 \subsection{Experimental design}

While the URE project goals were generally given, students felt they had significant autonomy to decide elements of experimental design. The specifics of what decisions students made varied greatly between groups, but there was extensive dialogue regarding the overall sense of autonomy and independence:

\begin{mybox}
{\noindent \emph{S1}: ``We get to choose how we want to accomplish different things. ... they could tell us to simulate something on Matlab or go build something on the [scanning tunneling microscope], ... they don't direct us that closely - but they do. It's not like we're choosing what we want to research in the field so much. Which, I feel like I wouldn't be able to do."

\noindent \emph{NH}: ``So the question is given to you, but the approach is up to?" 

\noindent \emph{S1}: ``Yeah, and we can ask for advice and stuff, but it's been a good balance, I think."

\noindent \emph{S2}: ``Yeah, I feel like, primarily, we need them to tell us what to do because we're not at a level where we can actually understand enough to choose what we want to do. So at the most basic level of, like, simulate exactly what happens under these conditions, we choose line-by-line what code to write, but they're telling us what they want the program to do."
}
\end{mybox}

Only two students expressed the dissenting view that in their UREs they had very limited autonomy in the decision making process. In both cases, much of the decision making was done for them and so it was unclear why they were doing what they were doing. One of the students attributed this to a project whose goal was to replicate an existing process. 

A group of students discussed how the design space demonstrated that physics is much more creative than they had previously thought. Students drew comparisons between exploring many possible designs with trying different approaches to solving homework problems. This was contrasted, however, with the fact that having a single correct answer or a single way to solve a problem removed much of the creative space in coursework. The fact that there was no single correct answer in their URE projects also led students to value the process of carefully analyzing the relevant variables in the system to check for systematic errors (their language was to ``see if you're missing something"). 

The sense of autonomy was also evident in the design lab course. For example, students described having the opportunity to decide what they needed to do the next day. One group highlighted that developing their own procedure taught them that the procedure you take is the most important element of an experiment, rather than the result produced---a conclusion that was also discussed in the context of their URE.

In contrast, students explicitly discussed how the structured lab courses did not allow autonomy. They saw no decisions to be made; they simply had to follow the instructions provided. Students contrasted this to being able to ``fiddle around" in their UREs to make sense of equipment. Students also mentioned having to refer to equipment manuals in their URE, with one student explicitly contrasting that they would never do this in a lab course. Students described often not paying attention to the process of the lab experiment because of the sheer amount of information given. Their focus on the experimental designs was also limited by time, because each week often involved a new lab activity that must be completed in a single two hour session.

\subsection{Construction and testing of apparatus (or code)}

In almost every interview, the majority of time was spent discussing this cognitive task, and every interview group had something to say about it. Troubleshooting was a significant part of the students' URE, from searching for bugs in code or finding leaks in vacuum equipment. One student described how ``everything is troubleshooting" and everything breaks. Troubleshooting was both frustrating and rewarding, and often cited as an element where the most learning occurred. When asked about content students felt was needed to prepare them for their URE, troubleshooting was the most common response. Most students said that no single course taught them troubleshooting, but some said it was picked up along the way to some degree. We will first discuss elements of construction.

Building or purchasing equipment occupied a significant amount of students' time early in the summer. One student said that they found making things very rewarding while others described building equipment as not yet `doing science'. The amount of building (whether code or equipment) in students' projects varied from building new systems, to tweaking or optimizing an existing system. Some were shocked, but always very grateful, that they were able to build equipment or code themselves, independently, during their URE. 

In this area, students found that there was much more `engineering-like' work than they had expected. They enjoyed being able to tinker with equipment:

\begin{mybox}
{\noindent \emph{S1}: ``There's a lot of manual, random tweaking you do, which I guess makes sense. But I feel like when I'm in the lab, I'm more of a car mechanic than like a scientist with a pristine white lab coat. That's actually interestingly fun - like one of the most fun things I did in the lab was change the oil in the vacuum pump."

\noindent \emph{S2}: ``That's very true! I've learned a lot about, like, torque wrenches and ... I know so much about how screws are measured."

\noindent \emph{S1}: ``Yeah." 

\noindent \emph{S2}: ``Like, what a quarter-twenty screw is and, like, all that sort of thing."

\noindent \emph{S1}: ``I feel like I can fix my car now. I feel like I can figure that out."

}
\end{mybox}

Students were surprised to learn about the enterprise, outside of the university, involved in building and supplying equipment to physics researchers. For example, one student mentioned that this presented career paths of which they had been previously unaware. Another noted the variety of personnel involved in building an experimental set-up with relatively specific expertise. This message conflicted with that of their degree program: the fact that students need to excel in a variety of courses in a variety of areas suggested to them that they needed expertise in all areas. 

In using existing equipment, students discussed how there were no instructions to follow. Instead, the student had to figure out why each piece was where it was. Students contrasted this with their experience in the structured labs; they were given instructions on how to use the equipment, and felt that they did not learn what the equipment was or how it worked. Working with oscilloscopes was extensively referenced in this context. 

Students whose URE projects involved constructing computer programs expressed a desire for the physics curriculum to include more explicit emphasis on learning how to code. They highlighted, however, that the content in existing programming-focused courses (both provided by the computer science and physics departments) was often insufficient or irrelevant to the specific tasks they needed in their URE. 

While some students said they did not know how to check whether things were correct, other students discussed testing their designs and codes in a variety of ways in their UREs. One student discussed using order of magnitude calculations to check for mistakes early. They discussed the skills required to be able to think, qualitatively or intuitively, on your feet, and how they encountered that in their URE. Another student discussed the process of breaking the system into smaller pieces to test individual elements and narrow down the problem. One group discussed how this process was much more fixed or rigid than the design process, in that there were specific tasks that needed to be to carried out to test each part of the system. This was contrasted with structured lab courses, however, where one student described explicitly that they would \emph{not} collect data on the performance of their equipment.

While troubleshooting was a source of much learning, it was also a source of frustration. Students were disheartened if their equipment broke early in the project, because significant use of valuable time was spent fixing it (again, rather than `doing science'). Nonetheless, many students found this process to be very rewarding, expressing a sense of satisfaction when they finally figured out how to fix their problem:

\begin{mybox}
{\noindent S1: ``When you break a machine in the one way that the professor said, `Do not break the machine because they don't make spare parts for this thing anymore.' But then you manage to fix it anyways and then the thing starts working again, that's good... Overcoming obstacles."

\noindent S2: ``I completely agree with that. Yesterday I was struggling all day long with how to fit this one graph a certain way and I was so upset and this morning I came in early and then it magically worked and I got it to work and I was so happy and it's carried me through the whole day."
}
\end{mybox}

This feeling of reward and satisfaction was even stronger if students had the independence to solve the problem themselves. Students described having to learn that one could try small modifications first, compared with their initial instinct to start over from the beginning when things went wrong. One student described eventually developing the courage to say ``I have no clue what's going on, but I can probably figure out why it's not working." Others said they learned not to get ``stressed about getting things perfect the first time." 

Regarding the iterative nature of experimentation, one student described how troubleshooting surprisingly led them to try to do things better. Another student described abandoning a project, instead of iterating to improve, because results were not promising enough to continue. A student highlighted the need for patience and that sometimes it was important to leave, go have lunch, and come back. They appreciated that research gave them time to step back and make sense of things (to be compared with structured lab courses). Iteration also came in the form of cycles of feedback and revisions with a professor or mentor. One student mentioned that, once they achieved data that suggested that the system was were working properly, the mentor would help figure out what to do next. 

Students noted that the independence and control over the project in the design lab courses made troubleshooting and flexibility necessary and much like their URE. Students also described having to suggest and find ways to improve their experiment. 

There was much less discussion about finally collecting data in the URE. Many students did not quite reach this stage until the very end of their project, since building and troubleshooting took so much time. Students who had functioning systems described the process of collecting data to be very rewarding. 

As suggested, students' descriptions of the structured lab courses paint a different picture. They described many of the structured lab experiments as having ``instant gratification" with results that came out as expected, with no need for iteration or reflection. Several interview groups highlighted a particular experiment in the introductory structured lab course when discussing troubleshooting: a Millikan oil drop experiment. Students described how, in this experiment, many students struggled to collect meaningful data. Students expressed frustration, however, at the contrast between this situation and their URE. Students said they did not have sufficient time or resources in the lab experiment to make sense of their messy data, identify all the sources of error, or figure out how to resolve them. From their URE, they recognized the importance of taking a step back from the activity, as well as focusing on the process instead of obtaining a specific result. They felt that the lack of time with each experiment created a missed opportunity. 

Another reason students did not engage in troubleshooting or spend time making sense of equipment in structured labs had to do with motivation. Students described the limitations when they were performing the experiment for a class assignment, rather than an authentic purpose. For example, a student said that, in research, they took care with their process because everything is `real' in research: systems break and it is their job to get them working again in order to move forward. In lab courses, you just need the machine to ``spit out numbers that agree with your calculations." We should note here that the notion of tackling cutting edge research questions was not mentioned in this use of authentic or real experimentation. That is, it was the process of the experimentation that they found to be particularly inauthentic, not the subject.

\subsection{Analyzing data}

Most students had not collected data until the end of the summer, and then only to a very limited extent. Regardless, there was extensive discussion about the analysis process in their URE. Many students spoke about data analysis in the context of the required statistics and programming knowledge. There was disagreement about whether targeted courses were helpful or sufficient preparation for the programming and statistics needs in their UREs. Others described the process of modeling data, especially identifying and understanding approximations and assumptions. There was also limited discussion about analysis related to producing graphs and using data to make statistical or qualitative arguments.

In the structured lab courses, the Millikan oil drop experiment was raised as one that required students to evaluate sources of error and uncertainty, though with limited time to act on that evaluation. Regarding the design lab courses, students described engaging in modeling and analyzing their data. One student described seeing little emphasis on the modeling process in their coursework, beyond just applying equations to solve problems.

\subsection{Evaluating results and analyzing implications}

Once again, there was little discussion about evaluating results because few students had obtained results. Students did describe, however, the questions that needed to be asked when interpreting results. In relation to checking results that come out the same as or different than expected, one student described not knowing how to check whether the data were correct. Another student described the exciting process of interpreting results with their mentor, asking questions such as ``are the data what we expected? Is something wrong or is it something new?" One student described their surprise at the subjectivity involved in making sense of results, recognizing that there may be multiple reasonable explanations for data. This was contrasted to coursework with a single correct answer. 

The notion of checking results came up more frequently with regards to lab courses. As mentioned earlier, students described that their structured lab course experiments typically involved obtaining a known or given value. A student described that, it was okay if their result came out differently than expected, and that they would just submit the wrong result. Other students described asking the TAs to tell them what result they were supposed to get and how they should get it, so that they could finish. Other students described getting frustrated when the experiment did not work, leading to temptations to ``massage data" to obtain the specific, known value. In their URE, these students had described learning to value data and their process because it might tell you what went wrong. In structured labs, students did not have time or incentive to figure that out:

\begin{mybox}
{\noindent S1: ``I mean, it's like, very much like you follow these - [the design lab course] is kind of different - but it's like you follow these steps and then here's, `Today we're learning about the charge of the electron.' And then you run this little experiment and you get a number that's, like, kind of close to right. Maybe it's, like, a multiple of 10..."

\noindent S2 (Interrupting): ``No no - when you do [that experiment] and you're, just like, no where close! I think I wrote in my lab report that `It is very disturbing that so many people have previously measured the electron charge incorrectly. We can clearly see from this one that it should be equal to that.'"
}
\end{mybox}

Discussions of the design lab course reflected the satisfaction, also described in the context of their UREs, that any result was a result. In this course, students also described being guided to evaluate what they can and cannot conclude from their data. 

\subsection{Presenting the work}

Regarding presentations, students primarily discussed the final poster presentation, which was a required part of the URE program. Students expressed that the fixed date of the poster presentation applied pressure to obtain presentable results. Presenting their work at group meetings was also raised as motivating them to make more progress. The notion of publishing did arise, but students did not expect to obtain a publication after only a summer of work. Some students mentioned learning about the process of publishing and some expressed surprise by the associated politics and pressure. 

Group meetings were also highlighted as helping students see the broader context of their and others' work. This was mentioned as being helpful to students who sometimes felt lost in the details of their URE project. They contrasted that this broader perspective was even harder to achieve in coursework. 

Students explicitly described \emph{not} presenting results of their experiments from structured lab courses. One student did mention that labs could provide some practice for writing reports on experiments, but that the lab courses they took did not. A couple of students noted that the design lab course helped them communicate their work and progress, through the use of lab notebooks. 

\begin{table*}
\caption{Broad summary of elements that are and are not included in undergraduate research, design lab courses, or structured lab courses.}\label{tab:summary}
\begin{center}
\begin{tabular}{lccc}
\hline
\hline
Cognitive Task Element & Research & Design lab courses & Structured lab courses \\
\hline
Establishing research goals & No & Yes & No \\
Define criteria for suitable evidence & Partially & Partially & No \\
Determine feasibility of experiment & Partially & No & No \\
Experimental design & Yes & Yes & No \\
Construction \& testing of apparatus & Yes & Yes & No (except collect data) \\
Analyzing data & Yes & Yes & Yes \\
Evaluating results \& Analyzing implications & No & No & No \\
Presenting the work & Yes & No & No \\
\hline
\end{tabular}
\end{center}
\end{table*}%

\section{Discussion and conclusions}

In this paper, we used focus group interviews with summer URE students in a physics department to first evaluate the cognitive tasks students engaged in during their UREs, with comparisons to students' experiences in typical coursework and lab coursework (Table \ref{tab:summary}). Students generally engaged with most of the cognitive tasks required in physics research during their UREs and design lab courses. The exception was initial goal setting and evaluation of feasibility in UREs. In coursework and structured lab courses, however, there was little engagement in many of the EPCTA elements.

Our second research question regarded the barriers to engagement in these tasks in these different settings. The primary barrier to engaging in cognitive tasks in their URE, extracted from student comments, was time. The limited term of the URE restricted the scope of the project. The amount of time taken up by building and testing equipment took over the other aspects of the EPCTA. The second significant barrier was students' content knowledge. As highlighted in the EPCTA \cite{Wieman2015}, many of the tasks require an extensive understanding of the current state of knowledge in the field, as well as technical expertise. 

As a result of these barriers, initial goal setting, evaluating criteria, and evaluating feasibility were typically conducted by the research mentors before the students began their projects. This seems necessary for students to engage with the second half of the cognitive tasks and be able to present results at the end of the work period. Both of these barriers are logistically impractical to overcome. One potential partial remedy would be for the research mentor to explicitly expose the student to the steps and decisions that led to the research question. 

It is important to remember that not all undergraduate science majors, however, can engage in UREs. How well do our courses (especially lab courses) expose students to the EPCTA elements? The conclusion of this work is that they do not do this well, at least as they are experienced by these students in courses with relatively conventional designs.

While it is understandable that lecture courses would not engage students in cognitive tasks involved in physics experimentation, the lack of exposure in traditional lab courses where students are conducting experiments may seem surprising. Time, once again, becomes particularly problematic in lab courses in that carrying out an experiment and analyzing the data in just a few hours, often in a single week, makes the autonomy, reflection, and iteration that is fundamental to most elements of the EPCTA impossible. The structuring and framing of the activities, in addition, discourage engaging in many cognitive tasks. 

Our third research question was to identify ways to overcome the barriers in coursework in order to better prepare students for research experiences. From our data, two small manipulations might allow structured labs to better prepare students for the cognitive tasks involved in UREs: providing time for testing and troubleshooting equipment (for example, spreading experiment across multiple weeks) and placing emphasis on the quality of students' process rather than the product they obtain. It was clear that when the goal was to obtain a known result, this corrupted the process. It did not emerge from our work that it was important that the experimental outcome was producing new knowledge.

We see two prominent future research questions from these results. First, would engagement and preparation in EPCTA elements in early lab courses lead to more fulfilling engagement in URE research activities, further \emph{improving the benefits} of UREs? Because research on UREs have discussed relatively little about the role UREs play in developing these skills, perhaps lab courses should focus on explicit skills development beyond simple engagement. While an open design lab course, where students choose their research goal and design, seems to include all EPCTA elements, we must recognize the necessary constraints and scaffolding for students to engage productively in and learn from those activities. Further research should probe the quality of engagement beyond the quantity found here. 

Second, could engagement and preparation in EPCTA elements in early lab courses \emph{replace} UREs in terms of the non-cognitive benefits they afford? Understanding this relationship would also require understanding the role of authenticity and community in these experiences. Although not part of the data set of this paper, students discussed elsewhere in the interviews how the community and collaboration between the undergraduate students and their graduate student, post-doc, or professorial mentors were significant contributions to the rewarding experiences during their UREs. Their contributions to an existing body of knowledge, however, were mentioned much less so; mostly students noted how fundamentally small their contributions were. In addition, the authentic contributions were discussed much less and seemed less important to them than the authentic decision making that was involved. This was contrasted to finding correct answers through a single correct procedure as in the structured labs and coursework.

This work provides a foundational characterization of the types of skills and cognitive activities students engage in related to experimentation in physics. Beyond those listed above, we see new research questions that these data and results elicit regarding the role of lab courses and UREs in training STEM majors to understand and develop the tools and activities of an experimental scientist. 

\acknowledgements
We would like to thank the undergraduate volunteers who participated in this work, and Rick Pam and Lauren Tompkins for their support in recruiting these students.

\bibliography{UREPaper2}

\end{document}